\documentclass[conference]{IEEEtran}
\IEEEoverridecommandlockouts
\usepackage{cite}
\usepackage{amsmath,amssymb,amsfonts}
\usepackage{algorithmic}
\usepackage{graphicx}
\usepackage{textcomp}
\usepackage{xcolor}
\usepackage{url,hyperref,lineno,microtype,subcaption}
\usepackage{bm}
\usepackage{caption}
\usepackage{subcaption}
\usepackage{mathtools}

\def\BibTeX{{\rm B\kern-.05em{\sc i\kern-.025em b}\kern-.08em
    T\kern-.1667em\lower.7ex\hbox{E}\kern-.125emX}}
\begin{document}

\title{{Towards Semantic-Aware Transport Layer Protocols: \newline A Control Performance Perspective} \\
\thanks{The authors acknowledge the financial support by the Federal Ministry of Education and Research of Germany in the programme of “Souverän. Digital. Vernetzt.”. Joint project 6G-life, project identification number: 16KISK002}}

\author{\IEEEauthorblockN{Polina Kutsevol, Onur Ayan, Wolfgang Kellerer}
	\IEEEauthorblockA{Chair of Communication Networks\\ Technical University of Munich, Germany\\
		Email: \{polina.kutsevol, onur.ayan, wolfgang.kellerer\}@tum.de}}

\maketitle

\begin{abstract}
Networked control systems (NCSs) are an example of task-oriented communication systems, where the purpose of communication is real-time control of processes over a network. In the context of NCSs, with the processes sending their state measurements to the remote controllers, the deterioration of control performance due to the network congestion can be partly mitigated by shaping the traffic injected into the network at the transport layer (TL). In this work, we conduct an extensive performance evaluation of selected TL protocols and show that existing approaches from communication and control theories fail to deliver sufficient control performance in realistic network scenarios. Moreover, we propose a new semantic-aware TL policy, which uses the process state information to filter the most relevant updates and the network state information to prevent delays due to network congestion. The proposed mechanism is shown to outperform all the considered TL protocols with respect to control performance.

\end{abstract}

\section{Introduction}

\label{sec:intro}

\begin{figure}[t]
		\begin{center}
			\includegraphics[width=\columnwidth]{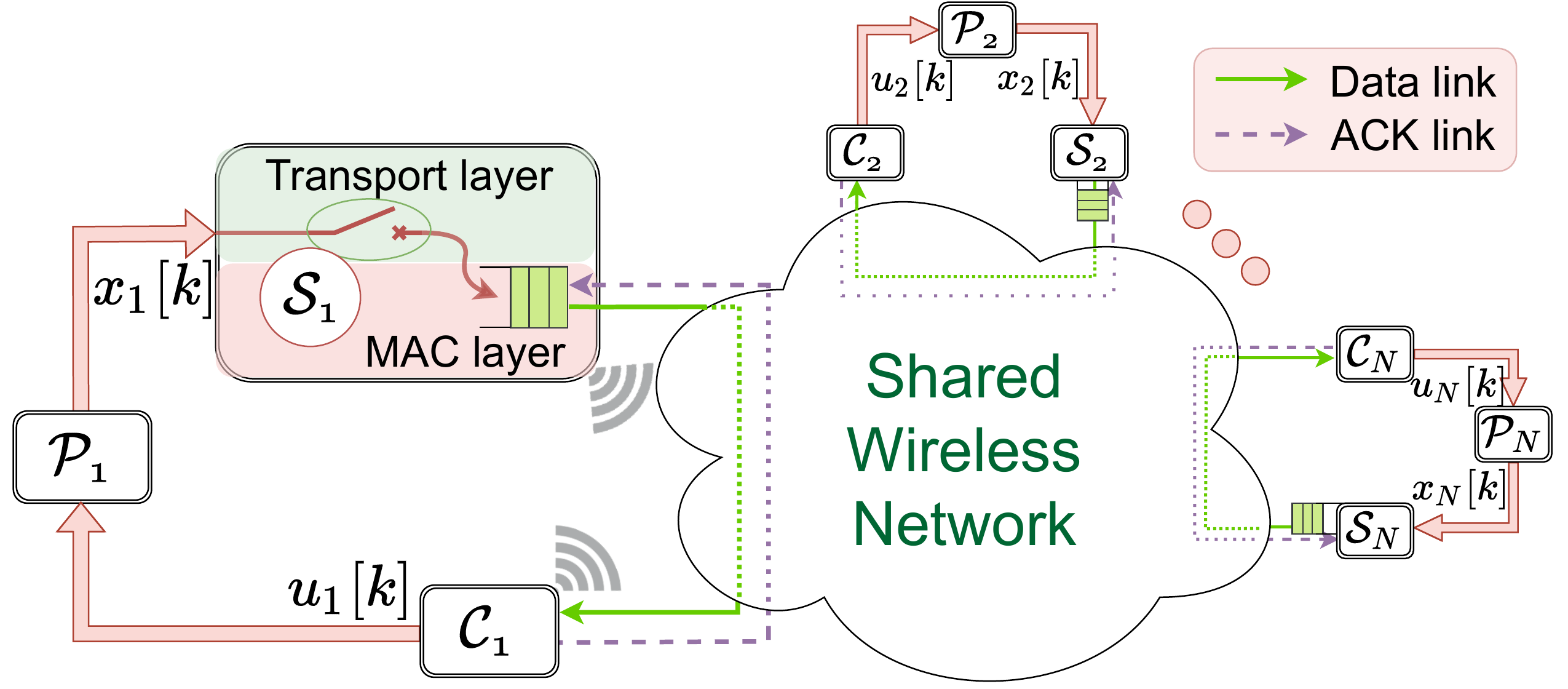}
		\end{center}

		\caption{The considered scenario with $N$ feedback control loops closed over the shared wireless network. In our implementation, we use various TL policies on top of FCFS MAC queue and CSMA/CA access scheme.}\label{fig:ncs_overview}

	\end{figure}

In the evolving post-Shannon 6G systems \cite{strinati20216g}, the perspective on the network is shifting from reliable transmission of bits to delivering heterogeneous services with respect to application-specific goals. A significant fraction of emerging applications stems from IoT. This application field of cyber-physical systems can be generalized based on their goal as networked control systems (NCSs). NCSs are feedback control loops closed over a communication network, containing the plant, the state of which should be managed by the controller based on updates regarding the plant status sampled by a sensor (see Fig.\ref{fig:ncs_overview}). The application goal of NCS can be formulated as keeping the plant at the desired state while minimizing the control efforts. 


For network design for NCSs, the main target is to develop cross-layer mechanisms for the network resources management with respect to the application goal. The approach of the modification of media layers, i.e., network, data link and physical, has not been favourable to the mainstream industry, since it limits the versatility of lower layers protocols and equipment \cite{uysal2021semantic}. Moreover, adopting cross-layer techniques involving these layers might be challenging during the deployment phase due to, e.g., the absence of dedicated hardware. Control-aware shaping of traffic at higher layers, e.g., transport, is another promising way of making the network operations compliant with the defined application goal. Thus, we focus on joint application and transport layer (TL) design, keeping the architecture transparent and flexible to lower layers.

One of the TL challenges in real-time monitoring and control over network is to determine the injection rate of the status update packets of the control process into the network. The authors of \cite{gautam2021comprehensive} indicate that both too low and too high sending rates result in unacceptable control performance due to insufficient data generation or increased delays caused by the network congestion. The operation of conventional TL policies as TCP and UDP is not tailored to the usage in the scope of time-critical applications \cite{beytur2020towards}. As an alternative, the work \cite{shreedhar2019age} presents a TL protocol adapting the sending rate for improved information freshness that is crucial for real-time remote monitoring and control. The information freshness is captured by age of information (AoI), which is defined as the elapsed time since the generation of the freshest packet available at the receiver. It is known that low update rates and high network delays contribute to the increase of AoI.

	Whereas minimization of AoI helps to find the balance between under-utilization and idleness and over-utilization and network congestion, AoI is agnostic to the content of the transmitted information and thus does not capture the most semantically relevant updates for the completion of the control task. To improve the performance metrics defined by the application, an efficient policy should make use of the available context specific to the application and the content of data to be sent when injecting packets into the network. We refer to it as ``task-orientation'' in our work.

    \subsection{Related Work}
    
	The authors of \cite{kountouris2021semantics, uysal2021semantic, strinati20216g} introduce the concept of semantic communication which puts the information significance as a basis for communication decisions, thus changing the paradigm of coping with the scarcity of network resources. 
	These studies suggest that the cross-layer design for data acquisition, communication and processing should take the semantic properties of shared information and communication purpose into account. Indeed, it is shown in \cite{maatouk2020age, wang2021value, sun2019sampling} that the consideration of the importance of the process state measurements allows to improve the performance of control and monitoring.

  The idea of exploiting application layer information is known in the control community. There, a common way to limit network resources utilization is  event-triggering (ET), i.e., sending the updates only upon the occurrence of an event indicating the urgency of a new transmission. 
  The works \cite{ramesh2016performance, molin2014price, vilgelm2016adaptive} investigate the performance of NCSs utilizing ET with multiple control loops, which share common network resources via contention-based network access mechanisms.

	\subsection{Our Contribution}
	
  Mentioned works make assumptions on the network which do not hold in many scenarios such as constant \cite{wang2021value} or negligible \cite{maatouk2020age,  ramesh2016performance, molin2014price} end-to-end transmission delays, instantaneous acknowledgements (ACKs) \cite{maatouk2020age, sun2019sampling} or absence of packet losses \cite{wang2021value, sun2019sampling}. In contrast to existing studies, we consider multiple control loops sharing one network prone to varying transmission delays and dropouts. We suppose that the control system is built on top of the existing network infrastructure, and can only be optimized at the software level, while the layers below TL can not be modified. We emulate a single hop network based on the carrier-sense multiple access with collision avoidance (CSMA/CA) medium access control (MAC) scheme, where the network routine is intertwined with the underlying control processes in real-time. With the implemented framework, we study different TL policies and their applicability in the context of NCSs. We show that conventional TL policies from the networking field fail to provide sufficient control performance. At the same time, ET schemes common in the control theory picking only the relevant updates for transmission are not able to cope with network congestion, because they are agnostic to the network state. Hence, we propose a distributed adaptive scheme that is both semantic- and network-aware and combines the ideas of ET and network congestion control to achieve enhanced control performance. 
	
	

	\section{System Model and Implementation}
	We consider $N$ feedback control loops closed over a shared communication network as depicted in Fig. \ref{fig:ncs_overview}.
	Sensors $\mathcal{S}_i, i \in \{1,...,N\}$ periodically measure system states of plants $\mathcal{P}_i$ and push them to the TL, where, according to the used TL protocol, some of these measurements are admitted to the sensor first come first serve (FCFS) MAC queue. Following the successful access by $\mathcal{S}_i$ and transmission over the shared network, the head of the queue packet is received by the respective controller $\mathcal{C}_i$. Based on the state updates, the controller determines the control input to be applied to drive the plant to a desired state. The controller-to-plant link is assumed to be ideal. We do not enforce synchronisation among loops.

	
	
	

	\subsection{Control Model}
	
	Each sub-system $i$ is modeled as a discrete-time, linear time-invariant (LTI) system with dynamics:
	\begin{equation}
	\bm{x}_i[k+1] = \bm{A}_i\bm{x}_{i}[k] + \bm{B}_i\bm{u}_i[k] + \bm{w}_i[k],
	\label{eq:dynamics}
	\end{equation}
	where $\bm{x}_i[k] \in \mathbb{R}^{n_i}$ denotes the state of the plant $\mathcal{P}_i$ at time step $k$. That is, $\bm{x}_i[k]$ is periodically updated with a constant sampling frequency and is assumed to be unchanged during any sampling period. The time-invariant state matrix $\bm{A}_i \in \mathbb{R}^{{n_i}\times {n_i}}$ defines the direct relationship between the current and the next state. $\bm{B}_i \in \mathbb{R}^{{n_i}\times {m_i}}$ is the input matrix, which determines how the next state is affected by the control input $\bm{u}_i[k] \in \mathbb{R}^{m_i}$. The process noise vector $\bm{w}_i[k]$ is i.i.d. Gaussian with covariance matrix $\bm{\Sigma}_i \in \mathbb{R}^{{n_i}_ \times {n_i}}$. We assume an initial state of $\bm{x}_i[0] = \bm{w}_i[0]$ and the setpoint value, i.e., the target value, to be at $\bm{0} \in \mathbb{R}^{n_i}, \forall i$. 
	
	The control policy to keep the plant at the desired state depends on the chosen cost function. We exploit a commonly used quadratic cost function $F_i$ referred to as linear-quadratic Gaussian (LQG) cost function which considers both the vicinity of the plant state to the target value and the control effort required to drive the system into equilibrium:
	\vspace*{-2mm}
	\begin{equation}
	F_i \triangleq \mathbb{E}\left[ \dfrac{1}{T} \sum_{k=0}^{T - 1} (\boldsymbol{x}_i[k])^T \boldsymbol{Q}_i \boldsymbol{x}_i[k] +  (\boldsymbol{u}_i[k])^T \boldsymbol{R}_i \boldsymbol{u}_i[k] \right],
	\label{eq:lqg}
	\end{equation} 
	where symmetric positive semi-definite matrices $\boldsymbol{Q}_i$ and $\boldsymbol{R}_i$ are the parameters for the customization of the cost function.
	
	The control law that minimizes the value of the cost defined in \eqref{eq:lqg} is the following:
	\vspace*{-2mm}
	\begin{equation}
	\label{eq:controllaw}
	\bm{u}_i[k] = - \bm{L^*}_i \bm{x}_i[k],
	\end{equation}
	where  feedback gain matrix $\bm{L^*}_i \in \mathbb{R}^{m\times n}$ is determined by the solution of the discrete algebraic Riccati equation as in \cite{wang2021value}. Note that the controller design happens prior to deployment, therefore, it is independent of the network performance. In this work, our goal is to adapt the network to a given controller instead of adapting the controller to a given network.
	
	In the presence of the network, any transmitted state might not be available to the controller instantaneously or can be lost. In order to compensate for these delays and losses, the controller estimates the plant state remotely based on the available past observations. That is, if $\nu_i(k)$ denotes the generation timestamp of the last available at the controller update $\bm{x}_i[\nu_i(k)]$, the state estimate $\hat{\bm{x}}_i[k]$ can be obtained similar to \cite{ayan2019age} as follows:
	\vspace*{-2mm}
	\begin{equation}
	\bm{\hat{x}}_i[k] = \bm{A}_i^{\Delta_i[k]} \bm{x}_i[\nu_i(k)] + \sum_{q =1}^{\Delta_i[k]} \bm{A}_i^{q - 1} \bm{B}_i \bm{u}_i[k - q],
	\label{eq:estimator}
	\end{equation}
	where $\Delta_i[k] = k - \nu_i(k)$ is a number of sampling periods elapsed since the generation of the last state update received by the controller, or age of the freshest information at the destination. Since only the estimation $\bm{\hat{x}}_i[k]$ is available at the controller, it uses it instead of the exact state $\bm{x}_i[k]$ when calculating the next control input in  \eqref{eq:controllaw}. The better state estimation $\bm{\hat{x}}_i[k]$, i.e., the smaller estimation error $\bm{\hat{x}}_i[k] - \bm{x}_i[k]$, leads to the control input being closer to the optimum, what reduces the resulting state deviation from the desired point and saves future control efforts. Thus, the minimization of the estimation error contributes to the improvement of the LQG cost in \eqref{eq:lqg}. Note that as shown in \cite{ayan2019age}, the decrease in $\Delta_i$ is expected to reduce the estimation error.

	\subsection{Network Model}
	The operation of each sensor can be viewed in two independent blocks. The sensor structure scheme is shown in Fig.\ref{fig:ncs_overview}. The \textit{transport layer} decides which packets to admit to the communication network. The \textit{MAC layer} is responsible for storing the admitted packets for future transmissions and managing the channel access in a distributed fashion.
	We assume a clear separation between TL and MAC layer. TL does not possess any control over packets that have already been forwarded to the lower layers. Moreover, each layer has its own ACK scheme which can only affect decisions within the corresponding layer. We allow modifications to the software-defined TL whereas the operation of the lower layers is fixed. With such an approach, we aim at simpler implementation and faster adoption in real-life industrial scenarios.

	The state measurements admitted by the TL protocol are pushed to FCFS MAC layer queues. Conditioned the queue is not empty, the sensor initiates the channel access procedure for the head of the queue packet following CSMA/CA protocol \cite{654749}. In case of successful transmission, the updates are received by corresponding controllers. If specified by the TL protocol, controllers respond with a TL ACK through the dedicated control channel. TL ACKs are the only feedback available at the TL at sensors. This implies that MAC specific information, such as queue size or queuing strategy, is unknown to the TL.

	\subsection{Implementation}
	In our implementation, plant states are sampled with the constant frequency according to \eqref{eq:dynamics} within parallel processes. Each measurement is sent via a UDP socket in a separate packet to a process we call \textit{artificial network} that imitates a CSMA/CA network in real-time. That is, the packets experience real-time delays caused by emulated queueing and channel access. In particular, for each head of the queue packet, the random back off counter is drawn, which decrements over time. If counter of a user reaches zero, its packet is delayed with the transmission latency. In case of a simultaneous medium access, all transmission attempts are unsuccessful, and a retransmission procedure is initiated unless the maximum retransmission count has been reached. If successful, the packet is sent via a UDP socket to the controller process. Here, the control input \eqref{eq:controllaw} is calculated and sent back to the plant process via the ideal link modelled as another UDP socket. Simultaneously, if applicable a TL ACK is sent back through the network.

	\section{Control Performance of Existing Transport Layer Protocols}
	\label{sec:existingTL}

	
	
	
	\subsection{Conventional Transport Layer Protocols}

	\subsubsection{UDP}
	
	
	UDP is a simple TL protocol which does not utilize any retransmission schemes, thus, does not guarantee reliable data transfer. In our scenario, the absence of retransmissions might be beneficial in terms of the control performance, since outdated measurements would not consume the network resources. However, using UDP means that all the sampled state updates are admitted to the network, which might result in large delays and losses due to congestion, what is detrimental for keeping the plant stable.
	
   Let us introduce the binary variable $\delta_i[k]$, which equals $1$ if the state  $\bm{x}_i[k]$ of the plant $\mathcal{P}_i$ at time $k$ is admitted to the network, and $0$ otherwise. In case of UDP, $\delta_i[k] = 1$, $\forall i$, $\forall k$.
	
	
	
	\subsubsection{TCP}
	
	
	
	
	Unlike UDP, TCP avoids network congestion by limiting the number of outstanding packets.\footnote{We set the maximum transmission unit (MTU) value as the update packet size.} In our implementation, the sent packet is considered as outstanding until the ACK for this packet is received or a timeout event occurs. We denote the number of outstanding packets for sensor $\mathcal{S}_i$ at time step $k$ as $n^{out}_i[k]$. TCP refrains from new transmissions if the number of outstanding packets reaches the congestion window value , i.e., $CW_i[k]$, which constantly adapts such that the network is stressed precisely to its limit, i.e.:
    \begin{equation*}
        \delta_i[k] = 
        \begin{dcases}
        1,& \text{if } n^{out}_i[k] < CW_i[k]\\
        0,              & \text{otherwise.}
        \end{dcases}
	\end{equation*}
	
	Note that TCP tolerates high delays, which makes it unsuitable for NCSs. Indeed, to fill the congestion window, a sensor can admit multiple consecutive measurements to the network, fostering their queueing within the MAC layer what potentially deteriorates the control performance. Moreover, as has already been mentioned, using TCP, the control performance suffers from the retransmissions of outdated packets.

	\subsection{Zero-Wait Policy}
	Zero-wait (ZW) policy implies sending a single new update only after the previous one is received by the monitor, i.e., $CW_i[k] = 1, \forall k, \forall i$.  Hence, long waiting times of the updates in the MAC queue are avoided. Smaller delays contribute to the freshness of data at the monitor and enhanced control performance. 
	Note that in our implementation of ZW, if the ACK timeout occurs for a certain packet, it is not retransmitted but cleared from the outstanding packets list and discarded.
	
	
	\subsection{Age Control Protocol}
    In the context of real-time applications, the network resource usage optimization is often performed by exploiting the metric of age. Indeed, the waiting time due to network congestion and resource contention can be limited by minimizing AoI at the controller. To our knowledge, the Age Control Protocol (ACP) \cite{shreedhar2019age} is the first and the only complete TL technique that is designed for real-time monitoring applications. ACP adapts the rate of sending updates over network, such that it is not too low, with the age growing significantly due to rare state actualization, and not too high, with the updates becoming stale due to networking delay. 
    
    The algorithm estimates the age at the monitor, assuming that TL ACKs are instant and setting the expected age to the age of the last ACKed update. More specifically, the algorithm checks how the age is affected by the change in the number of updates in the network and adjusts the sending rate, such that the age in the next epoch is expected to decrease. For instance, if the sending rate increase has led to an increase in AoI in the current epoch, ACP would consider the network as congested and decrease the sending rate, such that the network would be offloaded, new updates would get to the monitor faster and the age would decrease. As a result, sensors tend to send updates at the maximum rate that does not result in AoI increase at the monitor.  Note that although ACP uses such a property of information as freshness, it does not exploit its content, i.e., the state of the plant, when making decisions. 

	\subsection{Network-Unaware Event-Triggering}
	As discussed in Section \ref{sec:intro}, ET is a widely adopted technique to reduce the amount of transmission, hence offloading the network. Leveraging the results from the control theory on NCSs, we consider the ET technique to filter ``less relevant'' status update packets by their content. The ET mechanism defines an event as a significant change in the system state that indicates the necessity of updating the controller. Here, the decision on the significance is binary and realized through a threshold-based policy. That is, only upon the occurrence of a change in the system state that is greater than a certain pre-defined threshold $\lambda_i$, the new state measurement is injected into the network. Otherwise, currently sampled updates are discarded, since they are not classified as informative.\footnote{If a packet is discarded, it is never considered for possible admission in the future. This implies that a packet can only be queued within the MAC layer.}
	As a result, the decisions of ET obey the following rule:
    \begin{equation*}
	    \delta_i[k] = \begin{dcases}
        1,& \text{if } |\bm{x}_i[k]| \ge \lambda_i\\
        0,              & \text{otherwise.}
    \end{dcases}
	\end{equation*}
	
    The state awareness of ET makes its ideas resonate with the definition of semantic communications given in \ref{sec:intro}. Indeed, the deviation of the state from the equilibrium, i.e., its absolute value, carries the semantic importance of the level of the plant disturbance, which directly affects the completion of the application-defined goal of control and influences the cost function defined by the application. The exploitation of the state for network-related decisions can, thus, be beneficial for the improvement of control-defined metrics. However, since the current network status does not affect the traffic admitted to the MAC layer of the sensor, the ET is expected to create possible network congestion, which, in turn, contradicts the primary goals of the TL protocol.

		\begin{figure}[t]
	\begin{center}
	\includegraphics[width=\columnwidth]{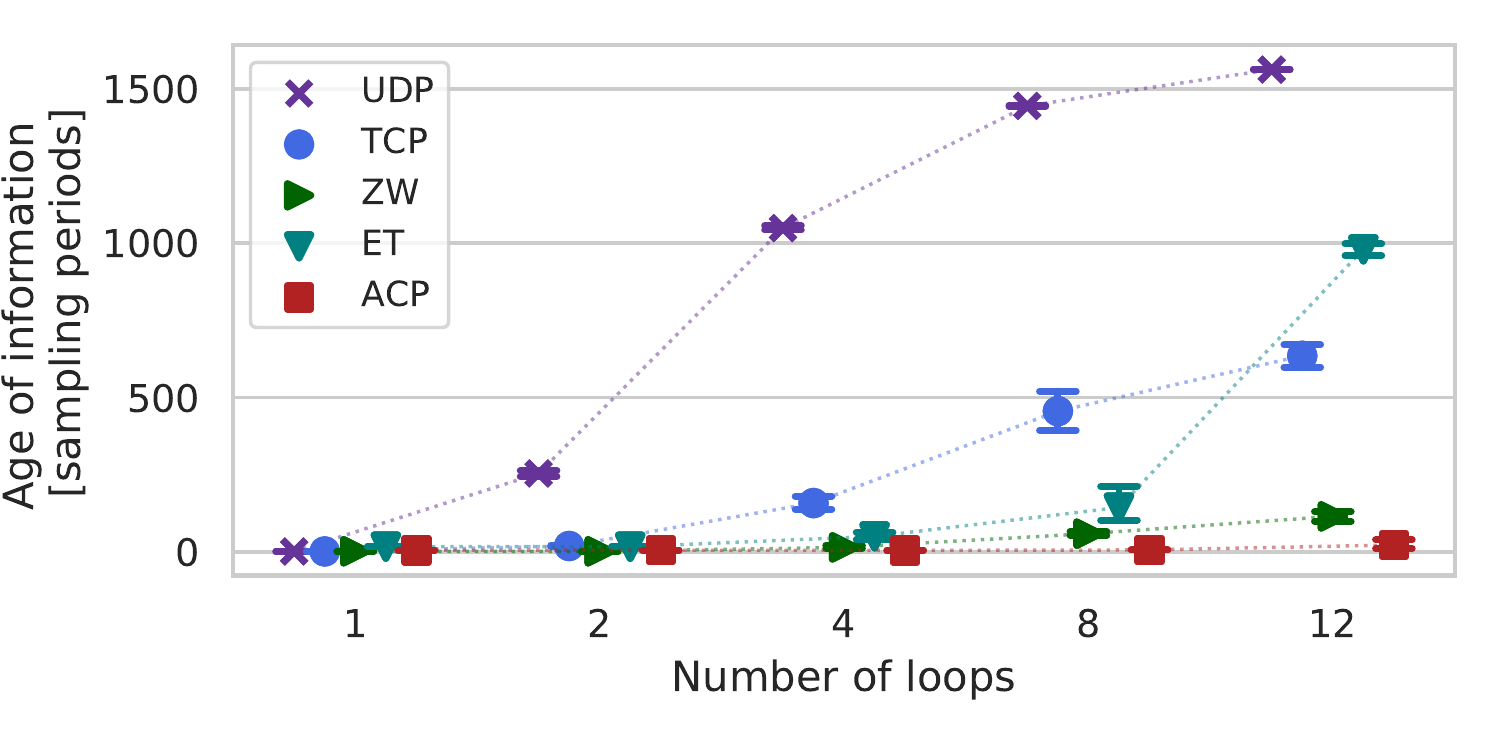}
	\end{center}
		\vspace*{-3mm}
	\caption{Average AoI and its 99\% confidence interval for UDP, TCP, zero-wait (ZW), event-triggering (ET) and age control protocol (ACP) policies. }
	\vspace*{-3mm}
	\label{fig:aoi_bad}
	\end{figure}
	
	\begin{figure}[t]
	\begin{center}
	\includegraphics[width=\columnwidth]{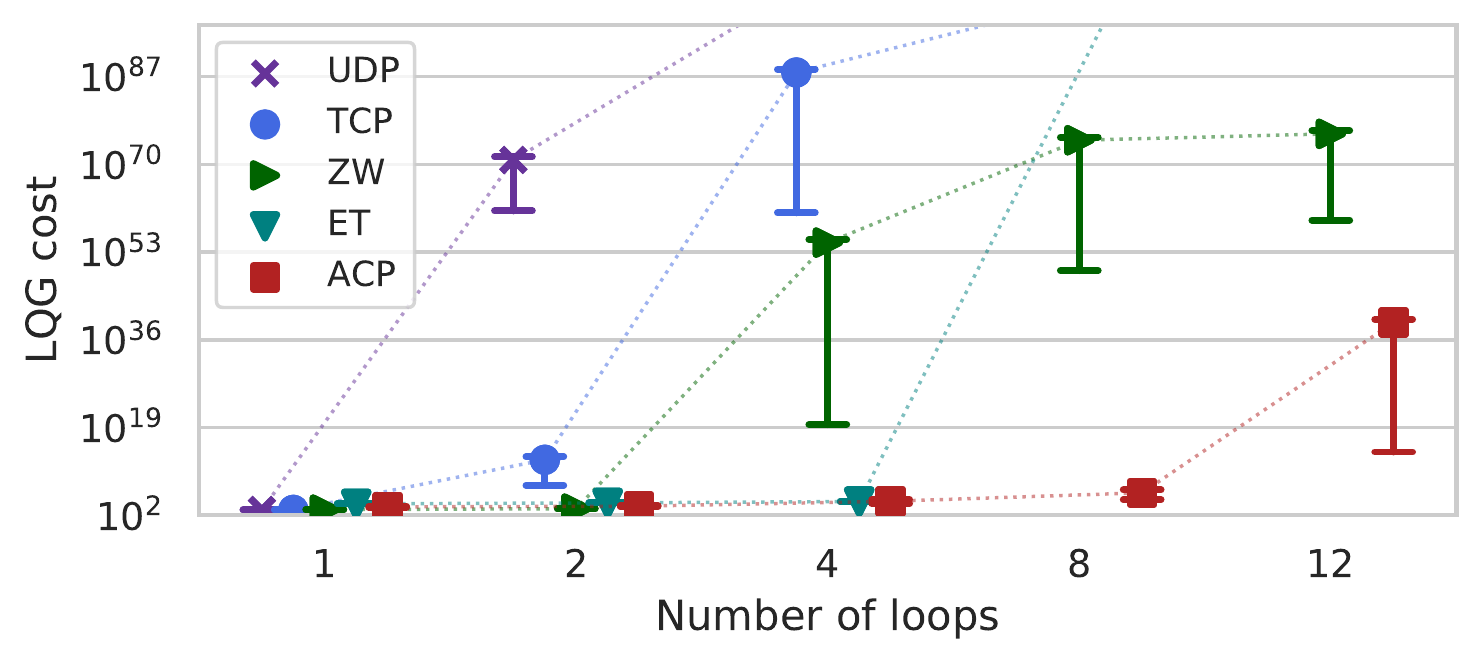}
	\end{center}
	\vspace*{-3mm}
	\caption{Average LQG cost and its 99\% confidence interval for UDP, TCP, ZW, ET and ACP policies. AoI-based ACP from \cite{shreedhar2019age} outperforms other techniques for $N=8$ and $N=12.$}
	\vspace*{-5mm}
	\label{fig:lqg_bad}
	\end{figure}

	\subsection{Numerical results for existing TL protocols}

	The performance of the described existing TL protocols is analyzed using our implemented NCSs framework. The following system parameters are used throughout all the measurements. In order to simplify the analysis, we use scalar control loops with state matrices $\bm{A}_i \in \{1.0, 1.1, 1.2\}$. We randomly assign state matrices values such that they are uniformly distributed for the scenario with the maximum number of loops, which is $12$. In particular, $\bm{A}_i = 1.0$ for $i\in \{4, 6, 8, 11\}$, $\bm{A}_i = 1.1$ for $i\in \{1, 2, 5, 12\}$ and $\bm{A}_i = 1.2$ for $i\in \{3, 7, 9, 10\}$. Moreover, the input and the noise covariance matrices are $\bm{B}_i=1, \forall i$ and $\bm{\Sigma}_i=1, \forall i$. For the LQG controller, we pick $\bm{Q}_i=100, \forall i$ and $\bm{R}_i=1, \forall i$, i.e., the growth of the plant error is penalized $100$ times more than the control effort. The duration of one measurement run is $30$s, whereas sampling periodicity is set to $10$ms, meaning that there are $3000$ $k$-steps. We perform $20$ measurement runs for every considered mechanism. 
	
	For network parameters, we use the values dictated by 802.15.4 \cite{7460875} often utilized in sensor networks. In particular, the backoff slot duration is $320$ms, and the maximum retransmission count is set to $7$\footnote{This is the default maximum retransmission count in \cite{1367266}}. The update transmission for $125$-bytes updates packets takes $4$ms, which corresponds to the typical data rate of 802.15.4 of $250$kbps. The transmission latency of the shorter TL ACK packet is set to $1$ms. 
	
	As performance metrics, we consider the average AoI and LQG cost in the network. The corresponding metrics for the single controller are calculated as averages for $k \in [1000, 3000]$ to exclude the transient effects at the beginning. As a result, the expression for the average AoI is the following:
	\vspace*{-2mm}
	\begin{equation}
	    \bar{\Delta} = \dfrac{1}{2001}\dfrac{1}{N}\sum_{k = 1000}^{3000} \sum_{i = 1}^{N} \Delta_i[k].
	    \label{eq:meanAoI}
	\end{equation} 
	The equation for the average LQG cost is derived from \eqref{eq:lqg} analogous to \eqref{eq:meanAoI}.

	AoI and LQG cost of UDP, TCP\footnote{In this work, we use TCP Tahoe version for the congestion control.}, ZW, ET and ACP are shown in Fig. \ref{fig:aoi_bad} and \ref{fig:lqg_bad}. Let us compare the performance of UDP, TCP and ZW protocols. Even though both AoI and LQG cost of TCP protocol are considerably better than of UDP, the usage of both schemes results in much higher age and control costs compared to the ZW policy. The reason for poor UDP performance lies in severe network congestion due to high access delay as a result of collisions. Indeed, if the network access time exceeds the packet arrival rate, which is sampling periodicity in the case of UDP, the FCFS MAC queue at sensors becomes a system bottleneck. TCP limits this congestion by pushing fewer updates to the network, however, large delays inherent to it and unnecessary retransmissions deteriorate the control performance. On the other hand, the reason for the better ZW performance is that it immediately reacts to the current network status, i.e., refrains from new transmissions when the network is busy serving the previous packet. 
	
	
	A clear disadvantage of ZW is that all users simultaneously try to access the network, which results in increased access delays. Therefore, ZW  falls behind ACP, especially for the higher number of loops. Indeed, the adaptation procedure of ACP learns the interplay between the sending rate and the AoI. Thus, the ACP forces the sensors to refrain from transmissions if it contributes to the decrease in AoI. As a result, the explicit consideration of age makes ACP the best performing TL policy among the selected state-unaware techniques w.r.t. information freshness. Facilitating the fresh updates being available at the controller for the state estimation, the ACP outperforms UDP, TCP and ZW protocols in control performance for our considered scenario, as it is shown in Fig. \ref{fig:lqg_bad}.

	Among the selected state-of-the-art (SotA) protocols, ET is the only state-aware technique. Although ET almost always outperforms TCP and UDP, it falls behind ZW and ACP. One clear disadvantage of conventional ET is that as the plant state reaches the threshold, all consecutive samples are likely to be admitted until the state is driven below the threshold, needlessly occupying network resources. This effect is visible in Fig. \ref{fig:lqg_bad} as the LQG cost diverges for ET already for $N = 8$. Thus, the potential benefit of content awareness is dominated by the shortcomings due to network unawareness. We can conclude that taking the network state into account is crucial for the realizations of real-time NCSs to obtain acceptable control performance in dense scenarios. Nevertheless, the consideration of the system state is expected to lead to a further performance increase if severe network congestion is avoided. Motivated by the standalone limitations of both approaches, we show how they can be combined in the context of NCSs.

	
		\begin{figure}[t]
	\begin{center}
	\includegraphics[width=\columnwidth]{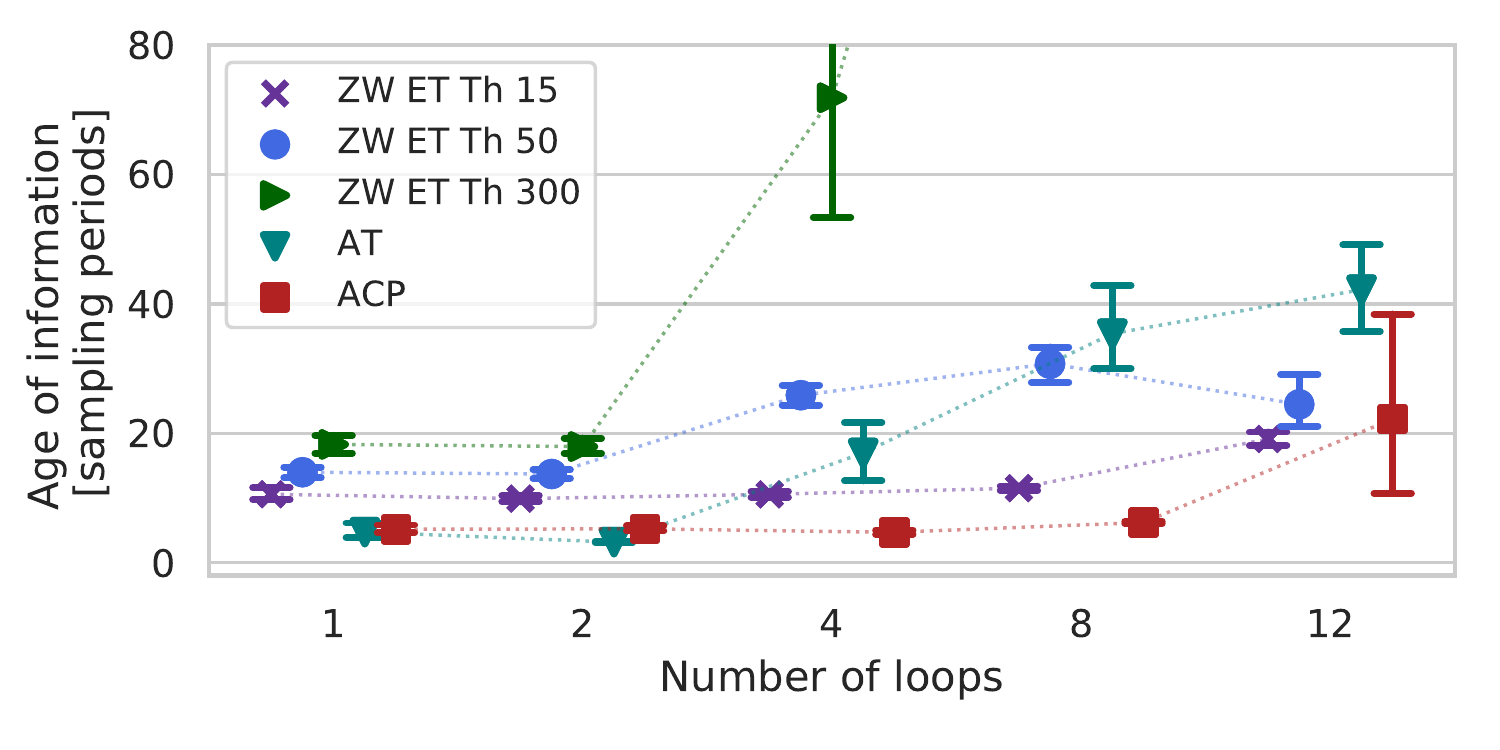}
	\end{center}
	\vspace*{-3mm}
	\caption{Average AoI and its 99\% confidence intervals for zero-wait event-triggering (ZW ET) with different thresholds, adaptive threshold (AT) and ACP.}
	\vspace*{-5mm}
	\label{fig:aoi_good}
	\end{figure}
	
	\begin{figure}[t]
	\begin{center}
	\includegraphics[width=\columnwidth]{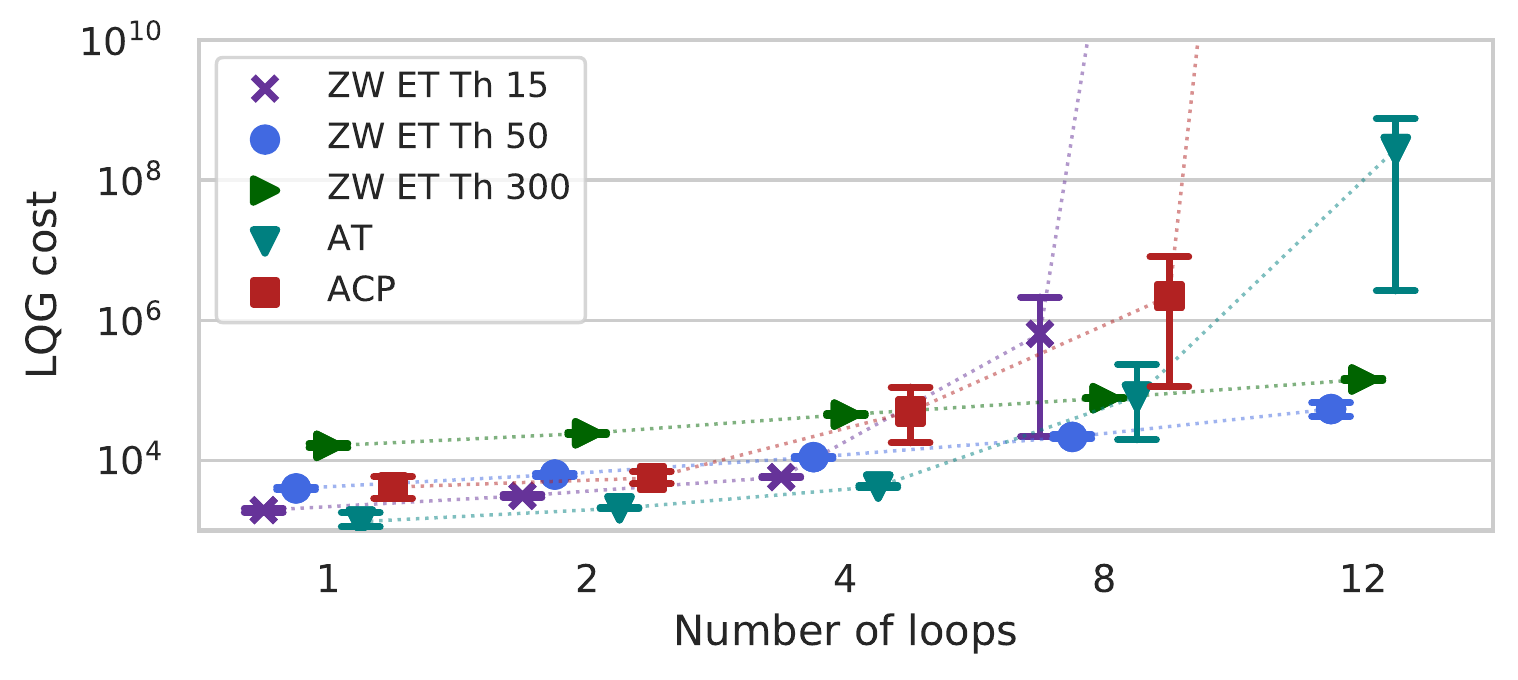}
	\end{center}
	\vspace*{-3mm}
	\caption{Average LQG cost and its 99\% confidence intervals for ZW ET with different thresholds, AT and ACP. Semantic- and network-aware techniques outperform ACP, which aims at improving information freshness.}
	\vspace*{-5mm}
	\label{fig:lqg_good}
	\end{figure}
	\section{Transport Layer Protocol for NCSs}
	\subsection{Zero-Wait Event-Triggering}

To prevent delays caused by bursty injections of the packets to the network as in the case of ET, we propose to combine conventional ET with the ZW mechanism, where the next packet is admitted to the network only after the previously sent packet has been served. Thus, in this work, we present our first semantic- and network-aware heuristic TL protocol for NCSs. In particular, in the proposed ZW ET policy, the packet is admitted to the network if both the current state is greater than the threshold and there are no outstanding packets, i.e.:
\vspace*{-1mm}
\begin{equation*}
	   \delta_i[k] = 
\begin{dcases}
    1,& \text{if } |\bm{x}_i[k]| \ge \lambda^i  \text{ and }  n_{out}^i[k] = 0\\
    0,              & \text{otherwise.}
\end{dcases}
\end{equation*}

	
	\subsubsection{Experimental Results}

	The numerical results for ZW ET for different threshold values are given in Fig. \ref{fig:aoi_good} and \ref{fig:lqg_good}. We also show ACP performance for comparison. Despite the superior age performance, ACP protocol falls behind ZW ET with an appropriate threshold in terms of LQG cost. The reason for that is the ability of ZW ET to limit the transmissions to only those that are relevant to the destination while keeping the queuing delays bounded thanks to the small congestion window. This proves that exploitation of the information semantics within updates allows to achieve better control performance compared to strategies optimized for age. Another indirect benefit of not allowing more than one outstanding packet in the network is the avoidance of bursty transmissions that are likely to carry less incremental update, although being currently considered relevant.
	
    Although ZW ET promises high potential due to its semantic awareness, it can be detrimental to control performance if an inappropriate threshold value for the considered scenario is chosen. For example, in the network with $N = 12$, $\lambda_i = 15, \forall i$ results in a significantly worse LQG cost than for a better threshold $\lambda_i = 50$, as it leads to too high sending rates for such a dense network. At the same time, the threshold value of $300$ is too conservative for $N = 1$, where the network could accommodate more frequent updates. 
	
	\subsection{Adaptive Threshold for ZW ET}
    In this section, we propose an add-on for the previously described ZW ET method, which is a reactive threshold adaptation scheme. The idea of the proposed mechanism is the following. Each sensor estimates the current level of congestion of the network locally. If it is increasing, the sensor raises its ET threshold, such that the criterium for an event is stricter. Otherwise, it concludes that there are available resources, thus the threshold can be decreased and more thorough control can be realized by admitting even slight state deviation messages.
	
	In particular, each sensor determines whether the last transmitted packet has been considered lost. If this is the case, i.e., its waiting time has exceeded the ACK timeout, witnessing network congestion, the current threshold is increased by $1.5$ times. If waiting times for all last $10$ packets have not exceeded the ACK timeout, that means the network is able to serve the current load and the threshold is decreased by $1.5$ times. 
	
	\subsubsection{Experimental Results}

	Fig.~\ref{fig:aoi_good} and \ref{fig:lqg_good} represent the performance of the proposed adaptive threshold (AT) scheme. It is shown to outperform the schemes with the fixed threshold for selected threshold values in terms of LQG cost for a low number of control loops, i.e., for $1$ - $4$ plants. This means that the network congestion is so low in these scenarios that the network would accommodate the updates from all the plants with a threshold lower than $15$, which is successfully captured by AT scheme. For a higher number of loops, e.g., for $12$, AT results in considerably better LQG cost than ZW ET with the "bad" threshold, e.g. $15$. However, its LQG cost is worse than ZW ET one with the "good" threshold of $50$. Note that AT scheme always outperforms state-unaware ACP.
	
	Thus, the proposed AT scheme represents a heuristic mechanism that can be used for the systems with varying number or dynamics of control loops or where brute force pre-computation of the optimal ET threshold is impossible. By using data semantics, AT achieves better control performance than state-unaware schemes. At the same time, the threshold adaptation prevents network under-utilization and severe congestion. Note that the threshold adaptation does not use semantic information, thus, this method does not achieve the best LQG cost performance observed for optimal fixed threshold in the ZW ET scheme.
	
	\section{Conclusion}

    The upcoming generation of wireless systems is expected to be tailored to task-specific requirements of services and applications. As shown in the SotA part, by the adoption of new metrics beyond delay and throughput, such as the freshness and relevance of information, the application layer performance can be significantly increased through cross-layer protocol design. In this work, we conduct an extensive study of various transport layer protocols from the literature in the context of control closed over network. Using a real-time measurement setup, we analyze the control performance of up to $12$ control systems sharing a common CSMA/CA network. In particular, we combine the core idea of event-triggering, which is a well-known technique from control theory to identify relevant information, with congestion control from communication theory to prevent large delays and losses. Our results suggest that the conventional transport layer protocols are inadequate when it comes to serving time-critical applications in a dense deployment scenario. Moreover, we show that exploitation of the information content allows to enhance the control performance compared to the existing protocols that only take information freshness into account, thus, state-agnostic.

\bibliography{biblio}
\bibliographystyle{IEEEtran}

\end{document}